\documentclass[letterpaper,11pt ]{article}
\usepackage{graphicx}% Include figure files
\usepackage{bm}% bold math
\usepackage[mathlines]{lineno}% Enable numbering of text and display math
\usepackage{amssymb,amsmath}
\usepackage{tabularx}
\usepackage{appendix}

\newcommand{\be}{\begin{equation}}
\newcommand{\ee}{\end{equation}}
\newcommand{\ben}{\begin{eqnarray}}
\newcommand{\een}{\end{eqnarray}}
\newcommand{\n}{\nonumber  }

\newcommand{\nd}{\noindent}
\begin{document}

%\preprint{Virial Ans\"atze}

\title{Virial ans\"atze for the Schr\"odinger Equation \\ with a symmetric strictly convex potential. Part II.}

\author{S. P. Flego\\
\\
\small{ Universidad Nacional de La Plata (UNLP), Facultad de Ingenier\'{\i}a,} \\
\small{GAMEFI-UIDET, (1900) La Plata, Buenos Aires, Argentina.}}
% \email{ flego@fisica.unlp.edu.ar}
%\date{\today}% It is always \today, today,
             %  but any date may be explicitly specified
	\maketitle

%\ead{flego@fisica.unlp.edu.ar \hskip 3mm (corresponding author)}

%%%%%%%%%%%%%%%%%%%%%%%%%%%%%%%%%%%%%%%%%%%%%%%%%%%%%%%%%%%%%%%%%%%%%%%%%%%%%%%%%%%%%%
\begin{abstract}
\nd Recently was introduced in the literature a procedure to obtain ans\"atze, 
free of parameters, for the eigenfunctions of the time-independent Schr\"odinger equation with symmetric convex potential. 
In the present work, we test this technique in regard to $x^{2\kappa}$-type potentials. We study the behavior of the ans\"atze regarding
  the degree of the potential and to the intervening coupling constant. 
Finally, we discuss how the results could be used to establish the upper bounds of the relative errors in situations where intervening polynomial potentials.
\end{abstract}
%\nd \pacs{05.45+b, 05.30-d}

\maketitle
\vspace{0.3cm}
\section{Introduction}
\nd Since only a few quantum-mechanical models admit of exact solutions,  approximations of diverse types constitute the tools  for to treat with the 
Schr\"odinger Equation (SE).  On the other hand, the Virial Theorem (VT) provides an extremely useful tackle for to study a quantum-mechanical system. In its non-relativistic quantum version, it is based on the SE. It relates the expectation value of the  kinetic energy of  the system to the expectation value of the directional derivative of the interaction potential that intervening in the Hamiltonian which describes the system  \cite{greiner}. 
Since it allows making conclusions about some interesting problems without solving the SE, since the 60s, hypervirial theorems  \cite{robi,castro} have been gainfully incorporated into the literature.  Furthermore, the subject was revisited in the Information Theory context, via the strong link between Fisher's Information Measure (FIM) and the SE.  In a nutshell, such connection is based upon the fact that a constrained minimization of the FIM leads to a SE-like equation [4-6]. This, in turn, implies intriguing relationships between various aspects of the SE, on the one hand, and the formalism of statistical mechanics as derived from the Jaynes's Maximum Entropy Principle, on the other hand [7 -14]. 
In particular, some fundamental consequences of the SE, such as the Hellmann-Feynman and Virial theorems, can be re-interpreted in terms of a special kind of reciprocity relations between relevant physical quantities similar to the ones exhibited by the thermodynamics' formalism \cite{HF-RR}.  This demonstrates that a Legendre transform structure underlies the non-relativistic SE \cite{HF-TV-RR}. 
As a direct consequence of this Legendre-symmetry that underlies the connection between the FIM and the SE, 
an ansatz for the ground state eigenfunction of a SE with convex even informational-potential was derived  \cite{pdf-1}. 
Then, in a quantum scenario,  as a direct consequence of the VT and symmetry considerations,  
a procedure to construct ans\"atze for the eigenfunctions of  the SE with symmetric convex  potential,  
was  developed \cite{ans-free-P1}.  The technique leads to the exact eigenfunctions of the harmonic oscillator, and it was successfully tested regarding the quartic anharmonic oscillator \cite{ans-free-P1}. 
	In the present communication,  the formalism introduced in \cite{ans-free-P1} is applied to $x^{2\kappa}$-type potentials. 
  We study the behavior of the ans\"atze in regard to the degree of the potential and to the intervening coupling constant. 
	Considering scaling properties, we show that 	the relative error $\epsilon_n$ in the  energy eigenvalue $E_n$   not depend on the coupling constant. 
 To illustrate the properties of the ansatz-solutions, we consider the quantum pure quartic oscillator and compare the results with those obtained by computational numerical calculation. Then we describe the principal characteristics that present oscillators of greater degree.
 Finally, we discuss how these results could be used to establish the upper bounds of the relative errors in situations where intervening polynomial potentials.
 \section{Preliminaries}
\subsection{The quantum scenario }
Let us consider the time-independent one-dimensional  Schr\"odinger wave equation (SE)  in dimensionless form,
 \ben \label{1} \left[-~\frac{1}{2}~\nabla_x^2~+~U(x)~\right]\psi_n(x)=E_n\psi_n(x)\:, \ \ \ \nabla_x^2~ \equiv ~\frac{d^2~}{d x^2}\:, \een  
\nd where  $ U(x)$ stands for a  time-independent symmetric convex real potential \cite{Nicu}.
 In this one-dimensional scenario, the {\it virial theorem } states  that \cite{greiner}
\ben \label{2} \left\langle - ~ \nabla_x^2 ~ \right\rangle_n  =  \left\langle x \: U\,'(x) \right\rangle_n ,  \een
 where  the expectation values are taken between stationary states of the Hamiltonian.
For one-dimensional scenarios,  $\psi_n$ is real \cite{richard},
then the VT can be written as
\ben \label{3} -\int_{-\infty}^{~\infty}{\psi_n(x) \nabla_x^2 \psi_n(x) \: dx}= ~\int _{-\infty}^{~\infty}{\psi_n^{\: 2}(x)~ x \:  U\,'(x)\:  dx} \:.\een
\subsection{Virial  ans\"atze for the eigenfunctions of the Schr\"odinger equation}
\nd Let U be a real, symmetric and strictly convex  potential which achieves its minimum at $x=\xi$.  
Then, from the VT and considering the symmetry property of the potential,  ans\"atze ${\chi} _n$  for the eigenfunctions ${\psi} _n$ of the SE are constructed  \cite{ans-free-P1}.  
They are given by
\ben \label{4}
\noindent \chi_n(x)=\varphi_n(x) \chi_v(x) \:, \een
where  the ${\chi}_v$-function is defined by
\ben \label{5}  {\chi}_{v}(x)=N~e^{-g(x)},\hspace{0.7cm}   \een 
with 
\ben \label{6}  g(x)=
\left\{ {\begin{array}{l}
	- \int{\sqrt{(x-\xi)\, U\,'(x)} dx} \hspace{0.8cm} if\hspace{0.2cm}~x < \xi\\
  + \int{\sqrt{(x-\xi)\,   U\,'(x)} dx}  \hspace{0.8cm} if\hspace{0.3cm}x \geq \xi\\
\end{array}} \right. , \een 
the constant $N$ determined by normalization condition and the functions $\{{\varphi}_n, n=0,1,2,\ldots\}$  chosen as a set of orthonormal polynomials  with weight function ${\sigma}(x)={\chi_v}^2(x)$,
\ben \label{7}
 \left\langle {{\varphi}_i {\varphi}_j } \right\rangle_{\sigma}  =  \int_{-\infty}^{~\infty}{{\varphi}_i(x) {\varphi}_j(x) ~{\sigma}(x)\;dx}=\delta_{ij} \:.
\een
The explicit form of the sequence $\{{\varphi}_n, n=0,1,2,\ldots\}$ can be established from the Gram-Schmidt orthonormalization  process \cite{szego} 
\ben \label{8}
{\varphi}_o(x)&=&1 ,\nonumber\\
{\varphi}_n(x)&=&
a_n\left[(x-\xi)^n-\sum_{k=0}^{n-1}{\left\langle  (x-\xi)^n   {\varphi}_k \right\rangle_{\sigma}}~{\varphi}_k(x)\right] \hspace{0.3cm} for ~ n\geq1 \:, \een
where 
\ben \label{9}
\left\langle {(x-\xi)^n~{\varphi}_k}  \right\rangle_{\sigma}  = \int_{-\infty}^{~\infty}{(x-\xi)^n ~ {\varphi}_k (x)~{\sigma}(x)~dx} \:,   \een
\nd
and the $a_n $ are constants determined by the normalization condition (\ref{7}).  Moreover, they can be expressed in terms of the Gram determinant \cite{szego}. 

\vspace{0.5cm}

\nd Once we have at our disposal the ans\"atze for the  eigenfunctions, we can obtain approximate energy eigenvalues,
\ben \label{10}
E_n  & \approx & E_n^{\it ansatz}= \left\langle \chi_{n} \left| H \right|  \chi_{n} \right\rangle=  \left\langle \chi_{n}\left|~-~\frac{1}{2}~\nabla_x^2~+~U(x)~ \right|  \chi_{n} \right\rangle . \een
Furthermore, they can be calculated using the virial theorem (\ref{2}), 
\ben \label{11}
 E_n^{\it ansatz}= \left\langle \chi_{n} \left| \frac{1}{2} ~(x-\xi)~U'(x)+U(x) \right| \chi_{n} \right\rangle . \een
\subsection{The Quartic Anharmonic Oscillator}
\nd The Schr\"odinger equation for a particle of unit mass in a quartic anharmonic potential reads
 \be \label{12}
\left[-~\frac{1}{2}~\nabla_x^2~
+~\frac{1}{2}~\omega^2~ x^2~+ ~\lambda ~x^4 \right]\psi_n~=
~ E_n~ \psi_n\:,   \hspace{1.cm} \lambda \geq 0 \:, \ee 
\nd where $\lambda$ is the coupling constant.

\nd The characteristics of the potential allow applying the technique cited above. 
The $\chi_v$-function (\ref{5}) is given by \cite{ans-free-P1}
\ben \label{13}
\chi_v (x) =N~exp\left[-{g(x)}\right]=N~exp\left\{\frac{\omega^3}{12\lambda}\left[1-
 \left(1+ \frac{4 \lambda}{\omega^2}  x^{2}\right)^{3/2}\right]\right\},\een
\nd where $N$ is determined by the normalization condition. The ans\"atze $\{\chi_n\}$ for the eigenfunctions $\{\psi_n\}$ are given by
 (\ref{4}) with $ \chi_v$ given by (\ref{13}) and,   being $\{ \varphi_n\}$ 
 a family of orthonormal real polynomials  associated  to the weight function $\sigma (x) =\chi_{v}^2(x)$, which can be obtained using  (\ref{8}).  

\nd  The approximate energy eigenvalues  (\ref{11}) are given by
\ben \label{14}
E_n \approx E^{\it ansatz}_n = \omega^2\left\langle  x^2 \right\rangle_n+3 \lambda \left\langle  x^4 \right\rangle_n \:, \een
where the expectation values are evaluated in respect to the ans\"atze $\chi_n$ for the eigenfunctions $\psi_n$.
\vspace{0.3cm}

\nd The ansatz curves for several sets of the coupling constant values can be observed in \cite{ans-free-P1}. There they were compared with the eigenfunctions curves numerically obtained. 
The curve associated with the ground state is bell-shaped.  It is observed that for small values of $\lambda$ $\left( \lambda \lesssim \omega^2/2\right)$ both curves overlap but,  as  the value of $\lambda$ grows, the width of the bell corresponding to the eigenfunction decreases a little faster than that of the curve corresponding to the ansatz. This difference propagates to the excited states. In the other hand, the approximate energy eigenvalues (\ref{14}) were compared with the numerical ones. Consistently the results show that, for small values of $\lambda$   the relative errors in the energy eigenvalues $\epsilon_n $ are small ($\epsilon_n \lesssim 1 \,\%$). When $\lambda$ increases, $\epsilon_n $ increases.  The question is, are these relative errors bounded?
We will return to this matter later.

\section{ Quantum oscillator with a  $x^{2\kappa}$-type potential.}
\nd In this section, we show the results obtained when dealing with monomial potentials. 
We are interested in the behavior of the ans\"atze, in regard to the degree of the potential and to the intervening coupling constant.
Besides the practical relevance of these potentials in connection with several areas of science, the results obtained here are going to be a useful tool to assess the applicability of the procedure in more complex situations.

\nd For reasons of clarity, in this section we will deal with even potentials.
The ansatz solutions  for symmetric potentials regarding $x = \xi$ can be obtained performing the translation transformation $x\rightarrow x-\xi$ in the obtained solutions for even potentials \cite{ans-free-P1}.

\subsection{Ans\"atze  for   $x^{2\kappa}$-type potential}
 \nd Consider the Schr\"odinger equation for a particle of unit mass, given by
\ben \label{AHO-1}
\left[-~\frac{1}{2}~\nabla_x^2~+~\lambda_{\kappa}~ x^{2\kappa}~\right] \psi_n^{(\kappa)}(x)~= ~ E_n^{(\kappa)}~\psi_n^{(\kappa)} (x)\:,  \hspace{1.cm}\kappa \in {\mathbf{N}}, \hspace{0.5cm} \lambda_{\kappa} > 0\:.\een
{\it The potential and its derivatives are given by}
\ben  \label{AHO-2}
U_{\kappa}(x)=   \lambda_{\kappa} x^{2\kappa}, \hspace{0.4cm}U'_{\kappa}(x) =  2\kappa \lambda_{\kappa} x^{2\kappa-1}, \hspace{0.4cm}
U''_{\kappa}(x)= 2\kappa (2\kappa-1) \lambda_{\kappa} x^{2\kappa-2}\:.\een
\nd Immediately, we observe that $U_{\kappa}$ is  a symmetric and strictly convex function\cite{Nicu}, with a unique minimum  at $x=0$,
\ben  \label{AHO-3}
\forall x \in \Re~, \hspace{0.4cm} U_{\kappa}(-x) = U_\kappa(x)~, \hspace{0.5cm}  U_\kappa'(0)=0  ~,\hspace{0.5cm}   U''_\kappa(x) \geq 0\:. \een
\vspace{0.1cm}
\nd Then, the characteristics of the potential allow applying the technique cited in the previous section.

\vspace{0.3cm}

\nd We start by constructing the virial $\chi_v$-function. Considering  (\ref{AHO-2}) we can write
\ben \label{AHO-4}
\int{\sqrt {x {U}'_\kappa(x)}dx}&=&\sqrt{2\kappa\lambda_{\kappa}} \int{|x|^\kappa dx}=
\left\{ \begin{array}{l}
- \frac{\sqrt{2\kappa \lambda_{\kappa}\;}}{(\kappa+1)} ~ |x|^{\kappa+1} \hspace{0.3cm}for ~x < 0 \\
+\frac{\sqrt{2\kappa \lambda_{\kappa}\;}}{(\kappa+1)}~ |x|^{\kappa+1}\hspace{0.3cm}for ~x \geq 0 \\
\end{array} \right. \:.\hspace*{0.2cm}\een
Substituting (\ref{AHO-4}) in (\ref{6}) we get
\ben \label{AHO-5}
{g}(x) &=&\hspace{0.2cm}~\frac{\sqrt{2\kappa\lambda_{\kappa}\;}}{\kappa+1}~|x|^{\kappa+1}~,\een
and the $\chi_v$-function (\ref{5}) is given by
\ben \label{AHO-6}
\chi_v^{(\kappa)} (x) =N_{\kappa}~exp{\left[-~\frac{\sqrt{2\kappa\lambda_{\kappa}\;}}{\kappa+1}~|x|^{\kappa+1}\right]}.\een
\normalsize
\nd Enforcing the normalization condition, we obtain
\ben \label{AHO-7}
 N_{\kappa}=\frac{1}{\sqrt{I_o^{(\kappa)}}}\left[{\frac{2 \sqrt{2\kappa \lambda_{\kappa} }}{\kappa+1}}\right]^{1/[2(\kappa+1)]},\hspace{1.cm}
 I_o^{(\kappa)}=\int_{-\infty}^{~\infty}{e^{-|x|^{\kappa+1}}} dx\:.\een
The virial weight function is given by
\ben \label{AHO-8}
\sigma_{\kappa}(x)=\left[\chi_{v}^{(\kappa)}(x)\right]^2
=N_{\kappa}^2  ~ exp{\left[- \frac{2 \sqrt{2\kappa\lambda_{\kappa}\;}}{\kappa+1} |x|^{\kappa+1}\right]} .  \een

\nd The ansatz $\chi_n^{(\kappa)}$ (\ref{4})  for the eigenfunction $\psi_n^{(\kappa)}$  can be written as 
 \ben \label{AHO-9}
\chi_n^{(\kappa)}(x)=\varphi_n^{(\kappa)}(x)~\chi_v^{(\kappa)} (x)\:, \een
with the functions $\{ {\varphi}_n^{(\kappa)}, n=0,1,2, \ldots\}$  chosen as a set of orthonormal polynomials  with weight function 
${\sigma_{\kappa}}(x)$ (\ref{AHO-8}).
\nd Explicit expressions of them can be obtained using the Gram-Schmidt process (\ref{8}).  
\nd It is computationally advantageous to express ${{\varphi}}_n^{(\kappa)}$ in terms of lower-order orthogonal polynomials
 using the {\it three-term recurrence} \cite{szego,chih}. For the present case, we have
\ben \label{TT-1}
{\varphi}_{0}^{(\kappa)}&=&1\:,\n\\
{\varphi}_{1}^{(\kappa)}&=& \langle x^2 \rangle_{\sigma_{\kappa}}^{-1/2}~x\:,\\
{\varphi}_{n}^{(\kappa)}&=&\beta_{n}\left[x~{\varphi}_{n-1}-
\left\langle x ~ \varphi_{n-1}{\varphi}_{n-2}\right\rangle_{\sigma_{\kappa}}~{\varphi}_{n-2}\right]\:,\ \ \ for~\:n\geq2\:, \n
\een
where the $\beta_n$ coefficients are given by
\ben \label{TT-2}
\beta_{n}=\left[\langle x^2 {\varphi}_{n-1}^2\rangle_{\sigma_{\kappa}}~-\left\langle {x ~\varphi}_{n-1}{\varphi}_{n-2}\right\rangle_{\sigma_{\kappa}}^2~\right]^{-1/2}. \n
\een

%%%%%%%%%%%%%%%%%%%%%%%%%%%%%%%%%%%%%%%%%%%%%%%%%%%%%%%%%%%%%%%%
\vspace{0.2cm}

\nd  The approximate energy eigenvalues  (\ref{11}) are given by
\ben  \label{AHO-10}
E_n^{(\kappa)}=\left\langle  U_\kappa(x)+\frac{1}{2}~x~U'_\kappa(x)\right\rangle_n^{(\kappa)}=
(\kappa+1) ~\lambda_{\kappa}~\left\langle  x^{2\kappa}\right\rangle_n^{(\kappa)}\:, 
\een
where $\left\langle  ~\right\rangle_n^{(\kappa)}$ denoted that the expectation values are taken in respect to the ans\"atze $\chi_n^{(\kappa)}$. 

\subsection{Behavior of the ans\"atze with the system parameters}
The explicit polynomial expressions of the $\varphi_n^{(\kappa)}$-functions can be written as
\ben \label{AHO-11} 
\varphi_n^{(\kappa)}(x)&=&  \sum_{j=0}^{n}{\alpha_{nj}^{(\kappa)}~x^j}\:, \een
where the coefficients $\alpha_{nj}^{(\kappa)}$ can be calculated   from the Gram-Schmidt orthonormalization  process (\ref{10}, \ref{11}).
 Immediately,  one can distinguish between even and odd orthonormal polynomial, 
$ \alpha_{nj}^{(\kappa)}=0$  when the parity of n is different from the parity of j.
The not null coefficients can be expressed in terms of the ground state moments,
\ben \label{AHO-12}
 \alpha_{nj}^{(\kappa)}= \alpha_{nj}^{(\kappa)} \left(\mu_o^{(\kappa)}, \mu_{2}^{(\kappa)}, \ldots, \mu_{2n}^{(\kappa)} \right) , \een
where 
\ben  \label{AHO-13}
\mu_{2i}^{(\kappa)}&=&
 \left\langle   x^{2i} \right\rangle_{\sigma_{\kappa}} =\int_{-\infty}^{~\infty}{x^{2i} \sigma_{\kappa}(x)~dx}=
\left(\frac{\kappa+1}{2 \sqrt{2 \kappa \lambda_{\kappa}}}\right)^{2i/(\kappa+1)}J_{2i}^{(\kappa)}\:, 
\een
with 
\ben  \label{AHO-14}
 J_{2i}^{(\kappa)}=\frac{I_{2i}^{(\kappa)}}{I_o^{(\kappa)}}~,
 \hspace{1.cm}I_{2i}^{(\kappa)} \equiv\int_{-\infty}^{~\infty}{ x^{2i}~e^{-|x|^{\kappa + 1}} dx} \:.
\een

\nd From (\ref{AHO-9}) and (\ref{AHO-11}), the ansatz $\chi_n^{(\kappa)}$ for the eigenfunction $\psi_n^{(\kappa)}$ can be written as 
\ben \label{AHO-15} 
\chi_n^{(\kappa)}(x)&=&  \sum_{j=0}^{n}{\alpha_{nj}^{(\kappa)}~x^j} ~\chi_v^{(\kappa)}(x) \:,  \een
  and the approximate energy eigenvalues  (\ref{AHO-10}) can be expressed as 
\ben \label{AHO-16}
E_n^{(\kappa)} &=& (\kappa+1) ~\lambda_{\kappa}~\int_{-\infty}^{~\infty}{x^{2\kappa} [\varphi_n^{(\kappa)}(x)]^2\sigma_{\kappa}(x)~dx}=\n\\
&=&  (\kappa+1) ~\lambda_{\kappa}~ \sum_{i,j=0}^{n} { \alpha_{ni}^{(\kappa)} \alpha_{nj}^{(\kappa)} \int_{-\infty}^{~\infty} {x^{2\kappa}x^{i+j} 
\sigma_{\kappa}(x)~dx} }=\n\\
 &=& (\kappa+1) ~\lambda_{\kappa}~\sum_{i,j=0}^{n} {\alpha_{ni}^{(\kappa)} \alpha_{nj}^{(\kappa)} \left\langle   x^{i+j+2\kappa} \right\rangle_o} \:. \een
Therefore, the energy eigenvalue  $E_n^{(\kappa)}$ depends on $\lambda_{\kappa}$ and  the first $2(n+\kappa)$ ground state moments,
\ben \label{AHO-17}
E_n^{(\kappa)} &=&E_n^{(\kappa)}\left( \lambda_{\kappa}, \mu_o^{(\kappa)}, \mu_{2}^{(\kappa)}, \ldots, \mu_{2(n+\kappa)}^{(\kappa)} \right) .
\een

%%%%%%%%%%%%%%%%%%%%%%%%%%%%%%%%%%%%%%%%%%%%%%%%%%%%%%%%%%%
\subsubsection{ Scale transformation properties}

\nd We describe here  the changes in the ansatz-solutions under a scaling transformation when intervenes a $x^{2\kappa}$-potential. 
For clarity, in the expressions of this section, we  drop the subscripts or superscripts that indicate the degree of the potential.

\nd It is well known that, under the scaling transformation $ x \rightarrow v=\lambda^{1/[2(\kappa+1)]} x$,  the SE  (\ref{AHO-1}) takes the form
\ben \label{scal-1}
\left[-~\frac{1}{2}~\frac{d^2}{d {v}^{2}}+v^{2\kappa}~\right]{\psi}_n^{sc}(v) ~= ~ {\cal E}_n~ {\psi}_n^{sc}(v) .\een 
\nd The scaling transformation relation between  the  eigenfunctions  and  between the energy eigenvalues are given by (see Appendix) 
\ben \label{scal-2}
{\psi}_n(x)=\lambda^{1/[4(\kappa+1)]} ~\psi_n^{sc}(\lambda^{1/[2(\kappa+1)]}~x) , \hspace{1.cm}
 E_n=\lambda^{1/(\kappa+1)}{\cal E}_n\:. \een

\vspace{0.3cm}

\nd $\star$ {\it The  scaling transformation of the ans\"atze}

\vspace{0.3cm}

\nd Following the procedure of the last section, we obtain the virial function for the transformed SE (\ref{scal-1}),
\ben \label{scal-3}
\chi_v^{sc} (v) =N^{sc}~exp{\left[-~\frac{\sqrt{2\kappa\;}}{\kappa+1}~|v|^{\kappa+1}\right]}.\een
\nd Enforcing the normalization condition, we obtain
\ben \label{scal-4}
 N^{sc}=\frac{1}{\sqrt{I_o}}\left[{\frac{2 \sqrt{2\kappa }}{\kappa+1}}\right]^{1/[2(\kappa+1)]},\hspace{1.cm} I_o
=\int_{-\infty}^{~\infty}{e^{-|v|^{\kappa+1}}} dv \:. \een
The ansatz $\chi_n^{sc}$ for the eigenfunction $\psi_n^{sc}$  is given by
 \ben \label{scal-5}
\chi_n^{sc}(v)=\varphi_n^{sc}(v)~\chi_v^{sc} (v) \:, \een
where the orthonormal polynomials $\varphi_n^{sc}$ associated to the virial weight function $\sigma^{sc}=[\chi^{sc}]^2$ have the form
\ben \label{scal-6} 
\varphi_n^{sc}(v)&=&  \sum_{j=0}^{n}{A_{nj}~v^j} \:, \een
where the coefficients $A_{ij}$ can be calculated  using (\ref{8}) and (\ref{9}).
The not null coefficients can be expressed in terms of the ground state moments,
\ben \label{scal-7} 
 A_{nj}^{sc}= A_{nj}^{(sc)} \left(\mu_o^{sc}, \mu_{2}^{sc}, \cdots, \mu_{2n}^{sc} \right)\:, 
\een
where 
\ben  \label{scal-8}
 \mu_{2i}^{sc}&=& \left\langle   v^{2i} \right\rangle_o^{sc} =\int_{-\infty}^{~\infty}{v^{2i}\sigma^{sc}(v)~dv}=
\left(\frac{\kappa+1}{2 \sqrt{2 \kappa}}\right)^{2i/(\kappa+1)}J_{2i} \:, 
 \een
with  $ J_{2i}$ given by (\ref{AHO-14}).
\nd The relationship between the transformed moments (\ref{scal-8}) and the original ones (\ref{AHO-13}) is given by 
\ben  \label{scal-9}
 \left\langle   x^{2i} \right\rangle_{\sigma_{\kappa}} =
 \lambda_{\kappa}^{- i/(\kappa+1)}~ \left\langle   x^{2i} \right\rangle_{\sigma_{\kappa}}^{sc} \:. 
\een
Considering (\ref{8}) and  (\ref{scal-9}), with a bit of algebra we obtain the relation between the  expansion coefficients of the orthonormal polynomials $ \varphi_n$ and $ \varphi_n^{sc}$,
\ben \label{scal-10}
\alpha_{nj}~=A_{nj}~\lambda^{j/[2(\kappa+1)} \:. \een

\vspace{0.3cm}

\nd Taking the relationship between the coefficients $\alpha_{nj}$ and $A_{nj}$  (\ref{scal-10}) into account,
 the inverse scaling transformation $ v \rightarrow x=\lambda^{-1/[2(\kappa+1)]} v$ leads to
\ben \label{scal-11}
\chi_n^{sc}(\lambda^{1/[2(\kappa+1)]}~x)&=&  \varphi_n^{sc}(\lambda^{1/[2(\kappa+1)]}~x) \chi_v^{sc}(\lambda^{1/[2(\kappa+1)]}~x)=\n\\
&=&N^{sc}~\left( \sum_{j=0}^{n}{A_{nj}~\lambda^{j/[2(\kappa+1)]}x^j}\right) ~exp{\left[-~\frac{\sqrt{2\kappa \lambda\;}}{\kappa+1}~|x|^{\kappa+1}\right]} =\n\\
&=& N^{sc}\left( \sum_{j=0}^{n}{\alpha_{nj}~x^j}\right) exp{\left[- \frac{\sqrt{2\kappa \lambda\;}}{\kappa+1}~|x|^{\kappa+1}\right]} \:.  \een
Considering (\ref{AHO-6}),  (\ref{AHO-9}) and  (\ref{AHO-11}),  the above equation can be written as 
\ben \label{scal-12}
\chi_n^{sc}(\lambda^{1/[2(\kappa+1)]}~x)=\frac{N^{sc}}{N} \varphi_n(x) \chi_v(x) =\frac{N^{sc}}{N}  \chi_n(x)  . \een
From (\ref{AHO-7}) and (\ref{scal-4}) we have
\ben \label{scal-13} \frac{N}{N^{sc}}=\lambda^{1/[4(\kappa+1)]}\:.   \een
Finally, substituting (\ref{scal-13}) in (\ref{scal-12}) we get
\ben  \label{scal-14} {\chi}_n(x)&=&\lambda^{1/[4(\kappa+1)]} ~\chi_n^{sc}(\lambda^{1/[2(\kappa+1)]}~x) \:. \een
Therefore, taking (\ref{scal-2}) and (\ref{scal-14}) into account, we can assert that the ans\"atze satisfy the same scale transformation property that the eigenfunctions satisfy .

 \vspace{0.5cm}

\nd $\star$ {\it The  scaling transformation of the energy eigenvalues}

\vspace{0.3cm}

\nd  Taking (\ref{AHO-10}) into account, the approximate energy eigenvalues for the transformed SE (\ref{scal-1}) are given by
\ben \label{scal-15}
{\cal E}_n^{ans} = \left\langle  U_\kappa^{sc}(v)+\frac{1}{2}~v~[U^{sc}_\kappa]'(v)\right\rangle_n^{sc}
=(\kappa+1) ~\left\langle  v^{2\kappa}\right\rangle_n^{sc}\:, \een
where the expectation values $\left\langle  ~\right\rangle_n^{sc}$ are taken in respect to the ansatz $\chi_n^{sc}$ for the eigenfunction $\psi_n^{sc}$. Using (\ref{scal-5}) and (\ref{scal-6}), with $\sigma^{sc}=[\chi^{sc}]^2$, we have 
\ben \label{scal-16}
{\cal E}_n^{ans} &=& (\kappa+1) ~\int_{-\infty}^{~\infty}{v^{2\kappa} [\varphi_n^{sc}(v)]^2~\sigma^{sc}(v)~dv}=\n\\
 &=& (\kappa+1) ~\sum_{i,j=0}^{n} {A_{ni} A_{nj} \left\langle   v^{i+j+2\kappa} \right\rangle_o^{sc}}\:. \een
Then, considering (\ref{scal-9}) and (\ref{scal-10}),  we can write
\ben \label{scal-17}
{\cal E}_n^{ans} &=&  (\kappa+1) ~\lambda^{\kappa/(\kappa + 1)}
 ~\sum_{i,j=0}^{n} {\alpha_{ni} \alpha_{nj} \left\langle   x^{i+j+2\kappa} \right\rangle_o} \:. \een
Finally, taking (\ref{AHO-16}) into account, we obtain
\ben \label{scal-18} {\cal E}_n^{ans} &=&~\lambda^{-1/(\kappa + 1)}~E_n^{ans} \:. \een
Therefore, the approximate energy eigenvalues, calculate using the  ans\"atze for the eigenfunction of the SE, satisfy the same scale transformation property that the energy eigenvalues satisfy (\ref{scal-2}),
\ben \label{scal-19} E_n^{ans}=   \lambda^{1/(\kappa + 1)} ~ {\cal E}_n^{ans}\:. \een
%%%%%%%%%%%%%%%%%%%%%%%%%%%%%%%%%%%%%%%%%%%%%%%%%%%%%%%%%%%%%%%%%%%%%%%%%%%%%

\subsubsection{Dependence of the relative errors on the coupling constant}
\nd Due to the scaling properties of the theory, the relative error in the ansatz-solution satisfies the following relation
\ben \label{scal-20} \hspace*{2.4cm}
\epsilon_n^{\chi}(x) = \frac{\left| \chi_n(x) -\psi_n(x) \right| }{\psi_n(x)}= \frac{\left| {\chi}_n^{sc}(v)-{\psi}_n^{sc}(v)\right|}{{\psi}_n^{sc}(v)}=\epsilon_n^{\chi^{sc}}(v),\hspace{0.3cm} \left(v=\lambda^{1/[2(\kappa+1)]}x\right), \een
and  the relative error in the energy eigenvalues not depend on coupling constant,
\ben \label{scal-21}
\epsilon_n^{}=\frac{\left|  E_n^{ans}-E_n\right|}{E_n}= \frac{\left| {\cal E}_n^{ans}-{\cal E}_n\right|}{{\cal E}_n} \:. \een
\subsection{Numerical results}
\nd To evaluate the ansatz-solutions, we are going to compare them with the eigenfunctions, and corresponding energy eigenvalues, obtained by computational numerical calculation. We will deal below with the harmonic oscillator ($\kappa=1$), with the quartic quantum oscillator ($\kappa=2$) and we will discuss the main features for quantum oscillators with $\kappa>2$.

\vspace{0.3cm}
\nd $\star$ {\large \it {The Harmonic Oscillator}  } 

\vspace{0.3cm}

\nd The Schr\"odinger equation for a particle of unit mass in a harmonic potential is given by,
\ben \label{HO-1}
\left[-~\frac{1}{2}~\nabla_x^2~+~\frac{1}{2}~\omega^2~ x^2~\right]\psi_n~= ~ E_n~\psi_n \:.  \een
 We can observe that the potential  is a convex even function  with a unique minimum  at $x=0$.
It is a $x^{2\kappa}$-type potential with $\kappa = 1$. This case was treated in \cite{ans-free-P1} where been shown that the ans\"atze  match the exact eigenfunctions of the SE (\ref{HO-1}).

\vspace{0.3cm}

\nd $\star$ {\large \it The Pure Quartic Oscillator } 

\vspace{0.3cm}

\nd The Schr\"odinger equation for a particle of unit mass in a pure quartic potential is given by,
\be \label{AHO4-1}
\left[-~\frac{1}{2}~\nabla_x^2~+~\lambda~ x^4~\right]\psi_n~= ~ E_n~\psi_n \:.  \ee
\nd We can observe that the potential  is a convex even function with a unique minimum  at $x=0$.
It is a $x^{2\kappa}$-type potential with $\kappa = 2$.
From   (\ref{AHO-6}) and (\ref{AHO-7}) with $\kappa=2$, the $\chi_v$-function is given by
\ben \label{AHO4-2}
\chi_v (x) =\frac{1}{\sqrt{I_o}} \left( \frac{4}{3} \right)^{1/6}~{\lambda^{1/12}}~exp{\left[-~\frac{2\sqrt{\lambda\;}}{3}~|x|^{3}\right]}, 
\hspace{1.cm} I_o=\int_{-\infty}^{~\infty}{e^{-|x|^{3}} dx}\:.  \een
\normalsize
For clarity, in the above expression, we omit the $\kappa$-subscript and the $(\kappa)$-superscript. 
Then, the ans\"atze $\chi_n$ (\ref{AHO-9}) for the eigenfunctions $\psi_n$ can be written as 
\ben \label{AHO4-3} 
\chi_n(x)&=& \frac{1}{\sqrt{I_o}} \left( \frac{4}{3} \right)^{1/6}~{\lambda^{1/12}}~ ~\varphi_n(x)~exp{\left[-~\frac{2\sqrt{\lambda\;}}{3}~|x|^{3}\right]} \:, \een
with the functions $\{{\varphi}_n, n=0,1,2...\}$  chosen as a set of orthonormal polynomials  with weight function $\sigma=\chi_v^2$, 
where $\chi_v$ is given by  (\ref{AHO4-2}).
Explicit expressions of them can be obtained using the Gram-Schmidt process (\ref{8}).  Furthermore, them can be expressed 
 in terms of lower-order orthogonal polynomials
 using the {\it three-term recurrence} (\ref{TT-1}).

\vspace{0.5cm}

\nd In each graph of figure 1,  the first five eigenfunctions of the quartic oscillator for a given $\lambda$-value are represented. 
They were obtained by computational numerically calculation (Matslise program was used) and the corresponding curves are drawn with black dashed lines. The corresponding ans\"atze (\ref{AHO4-3}) are represented in the same graph. Solid lines represent their curves 
(black for $\chi_o$, red for $\chi_1$, blue for 		$\chi_2$, green for $\chi_3$ and coral for $\chi_4$).  The curve associated with the ground state is bell-shaped.  It is observed that for this state, for all values of $\lambda$ considered, the width of the bell corresponding to the ansatz is a little greater than that of the curve corresponding to the eigenfunction. Consequently, due to the normalization condition, the ansatz curve looks lower. This difference between the ground state curves propagates to the excited states.

\small
\begin{center}
 {\bf Figure 1. Eigenfunctions of the pure quartic oscillator for several $\lambda$-values. }

\includegraphics[width=15.50cm,height=14.0cm,angle=0]{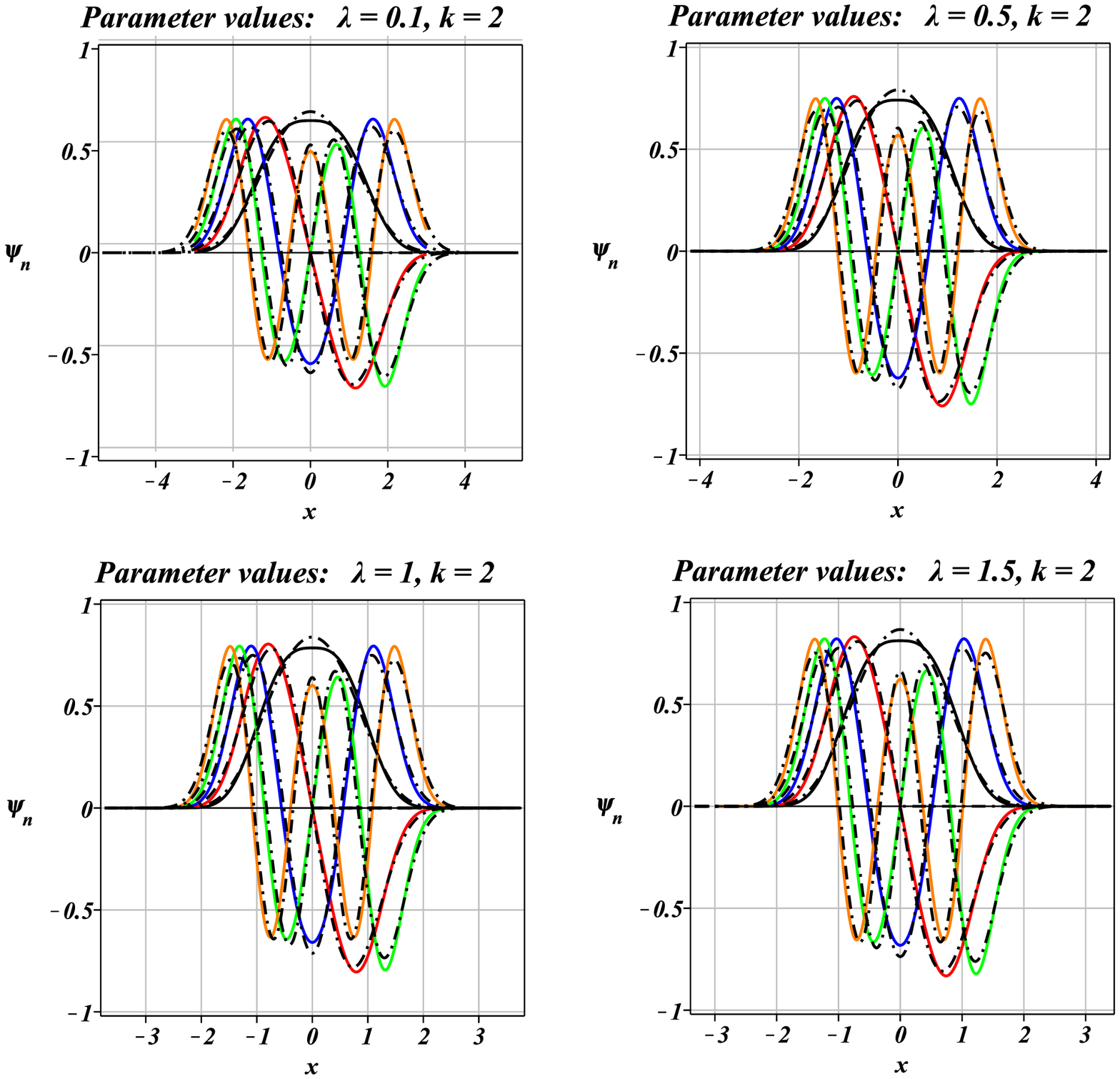}
\end{center}

\normalsize
\nd  From  (\ref{AHO-10}) with $\kappa=2$, the approximate energy eigenvalues   are given by
\ben \label{AHO4-6}
E_n \approx E_n^{ans}=3~\lambda ~\left\langle  x^4\right\rangle_n ,   \een
where the expectation values are evaluated in respect to the ansatz $\chi_n$ (\ref{AHO4-3}).

\vspace{0.3cm}

\nd In the tables 1 are tabulated the eigenvalues of energy corresponding to the eigenfunctions plotted in figure 1. In each table, in the first column are the principal quantum numbers. The values of the second column correspond to those energy eigenvalues that one finds in the literature,  obtained via a numerical approach to the SE. The values of the third column, correspond to the approximate energy eigenvalues obtained using (\ref{AHO4-6}). 
The fourth column displays the associated percentage relative error $\epsilon_n$, 
\ben \label{error}\epsilon_n  =\;\frac{|~E_n^{ans}- E_n|}{E_n}~100 \:.  \een
As we can observe, the percentage relative error for the ground state energy eigenvalue is of the order of $3.2\%$, 
and for the excited states, $n\geq 1$,  they are of the order of $1.7\%$. 
Furthermore, the results obtained show that the relative errors do not depend on the values of  $\lambda$, which was to be expected, according to (\ref{scal-21}).

\small
\begin{center}
\nd {\bf  Tables 1. Energy eigenvalues for the quartic  oscillator  for several  $\lambda$-values. \hspace*{0.7cm}}\\

\vspace{0.2cm}
\footnotesize

\begin{minipage}[t]{ 7.40cm}
\begin{tabular}{||c|r|r|r||}
\hline
\hline
\multicolumn{4}{|c|}{\bf  Parameter values: $k=2$,  $\lambda=0.1$}\\
\hline
$n$&$E_n\hspace{0.6cm}$&$E_n^{ans}\hspace{0.6cm}$&$\epsilon_n    \% \hspace{0.4cm} $\\
\hline
\hline
 0&0.31005176&0.31974622&3.12673 \\
 1&1.11103113&1.12975213&1.68501\\
 2&2.18005930&2.21667262&1.67946 \\
 3&3.40494424&3.46400460&1.73455 \\
 4&4.75498678&4.83502094&1.68316\\
 5&\hspace*{0.2cm} 6.21013792&6.31602813&1.70512 \\
\hline
\hline
\end{tabular}
\end{minipage}
\begin{minipage}[t]{ 7.40cm}
\begin{tabular}{||c|r|r|r||}
\hline
\hline
\multicolumn{4}{|c|}{\bf  Parameter values: $k=2$,  $\lambda=0.5$}\\
\hline
$n$&$E_n\hspace{0.6cm}$&$E_n^{ans}\hspace{0.6cm}$&$\epsilon_n    \% \hspace{0.4cm} $\\
\hline
\hline
   0&0.53018104&0.54675835&3.12673 \\
 1&1.89983651&1.93184898&1.68501\\
 2&3.72784897&3.79045685&1.67946\\
 3&5.82237276&5.92336456&1.73455\\
 4&8.13091301&8.26776951&1.68316\\
 5&10.61918647&10.80025618&1.70512\\
\hline
\hline
\end{tabular}
\end{minipage}

\vspace{0.2cm}

%%%%%%%%%%%%%%%%%%%%%%%%%%%%%%%%%%%%%%%%%%%%%%%%
\begin{minipage}[t]{ 7.40cm}
\begin{tabular}{||c|r|r|r||}
\hline
\hline
\multicolumn{4}{|c|}{\bf  Parameter values: $k=2$,  $\lambda=1.0$}\\
\hline
$n$&$E_n\hspace{0.6cm}$&$E_n^{ans}\hspace{0.6cm}$&$\epsilon_n    \% \hspace{0.4cm} $\\
\hline
\hline
   0&0.66798626&0.68887235&3.12673\\
 1&2.39364401&2.43397719&1.68501\\
 2&4.69679538&4.77567638&1.67946 \\
 3&7.33572999&7.46297170&1.73455\\
 4&10.24430846&10.41673681&1.68316\\
 5&13.37933656&13.60747014&1.70512 \\
\hline
\hline
\end{tabular}
\end{minipage}
\begin{minipage}[t]{ 7.40cm}
\begin{tabular}{||c|r|r|r||}
\hline
\hline
\multicolumn{4}{|c|}{\bf  Parameter values: $k=2$,  $\lambda=1.5$}\\
\hline
$n$&$E_n\hspace{0.6cm}$&$E_n^{ans}\hspace{0.6cm}$&$\epsilon_n    \% \hspace{0.4cm} $\\
\hline
\hline
 0&0.76465338&0.78856199&3.12673 \\
 1&2.74003839&2.78620836&1.68501 \\
 2&5.37648857&5.46678477&1.67946 \\
 3&8.39731462&8.54297000&1.73455 \\
 4&11.72680581&11.92418702&1.68316\\
 5&15.31551711&15.57666493&1.70512  \\
\hline
\hline
\end{tabular}
\end{minipage}
\end{center}

\vspace{0.3cm}

\normalsize

\nd $\star$ {\large \it The Pure Oscillator with $\kappa\geq 2$.} 

\vspace{0.2cm}

\nd For $x^{2\kappa}$-type potentials with $\kappa \geq 2$, the graphic representation of the ground state eigenfunction is a bell-shaped curve.
 It is found that for this state, as  the value of $\kappa$ grows, the width of the bell  decreases faster than that of the bell corresponding to the ansatz.   This discrepancy  propagates to the excited states.
Consequently, the relative errors in the energy eigenvalues increase as $\kappa$ increases. Once again, the results obtained show that the relative errors $\epsilon_n^{(\kappa)}$ do not depend on the coupling constant $\lambda_{\kappa}$,  according to (\ref{scal-21}).
\nd In table 2 are tabulated the percentage relative errors $\epsilon_n^{(\kappa)}$ for the ten first energy eigenvalues. Figure 2 shown  the percentage relative error  $\epsilon_n^{(\kappa)}$ as function of the principal quantum number $n$. Each curve correspond to a value of $\kappa$ 
(black for $\epsilon_n^{(1)}$, red for $\epsilon_n^{(2)}$, blue for 	$\epsilon_n^{(3)}$, green for $\epsilon_n^{(4)}$ and coral for $\epsilon_n^{(5)}$).
 As can be observed, $\epsilon_n^{(\kappa)}$ increases considerably as $\kappa$ increases. The maximum relative error occurs for the ground state energy eigenvalue. Then,  they decrease for $n \geq 1$, and  stay approximately constant for $n \gtrsim 3$,
\ben \epsilon_n^{(\kappa)} \approx 0.54 ~  \epsilon_o^{(\kappa)} \hspace{1.cm} for \hspace{0.3cm} n \gtrsim 3\:.  \een
 Furthermore, from (\ref{error}) we can write
\ben \label{error2}
E_n(\lambda) =\gamma_n^{\kappa} ~ E^{\it ansatz}_n(\lambda),\hspace{0.8cm}with \hspace{0.3cm} \gamma_n^{\kappa}=\left[1+\frac{\epsilon_n^{\kappa}}{100}\right]^{-1}\:. \een

\begin{minipage}{ 7.5 cm}
\footnotesize

\begin{tabular}{||c|r|r|r|r||}
\hline
\hline
\multicolumn{5}{|c|}{\bf  Table 2. Percentage relative error.  }  \\
\hline
$n$&$\epsilon_n^{(2)} $  \hspace{0.1cm} &$\epsilon_n^{(3)}$  \hspace{0.1cm} &$\epsilon_n^{(4)}$   \hspace{0.1cm}  & $\epsilon_n^{(5)}$   \hspace{0.1cm} \\
\hline
\hline
0&~ 3.1267&~ 9.8999&18.2593&27.4044\\
1&1.6850&5.7698&11.1600&17.2945\\
2&1.6795&5.2952&9.6670&14.3917\\
3&1.7345&5.5703&10.1473&14.9195\\
4&1.6831&5.4407&10.0471&14.9422\\
5&1.7051&5.4551&10.0049&14.8510\\
6&1.6901&5.4407&9.9983&14.8357\\
7&1.6967&5.4379&9.9839&14.8169\\
8&1.6915&5.4343&9.9784&14.8048\\
9&1.6939&5.4322&9.9729&14.7967\\
10&1.6917&5.4306&9.9696&14.7905\\
\hline
\hline
\end{tabular}

\end{minipage}
\begin{minipage}{ 7.5cm}
\vspace{0.2cm}

\begin{flushright}
\includegraphics[width=6.5cm,height=5.8cm,angle=0]{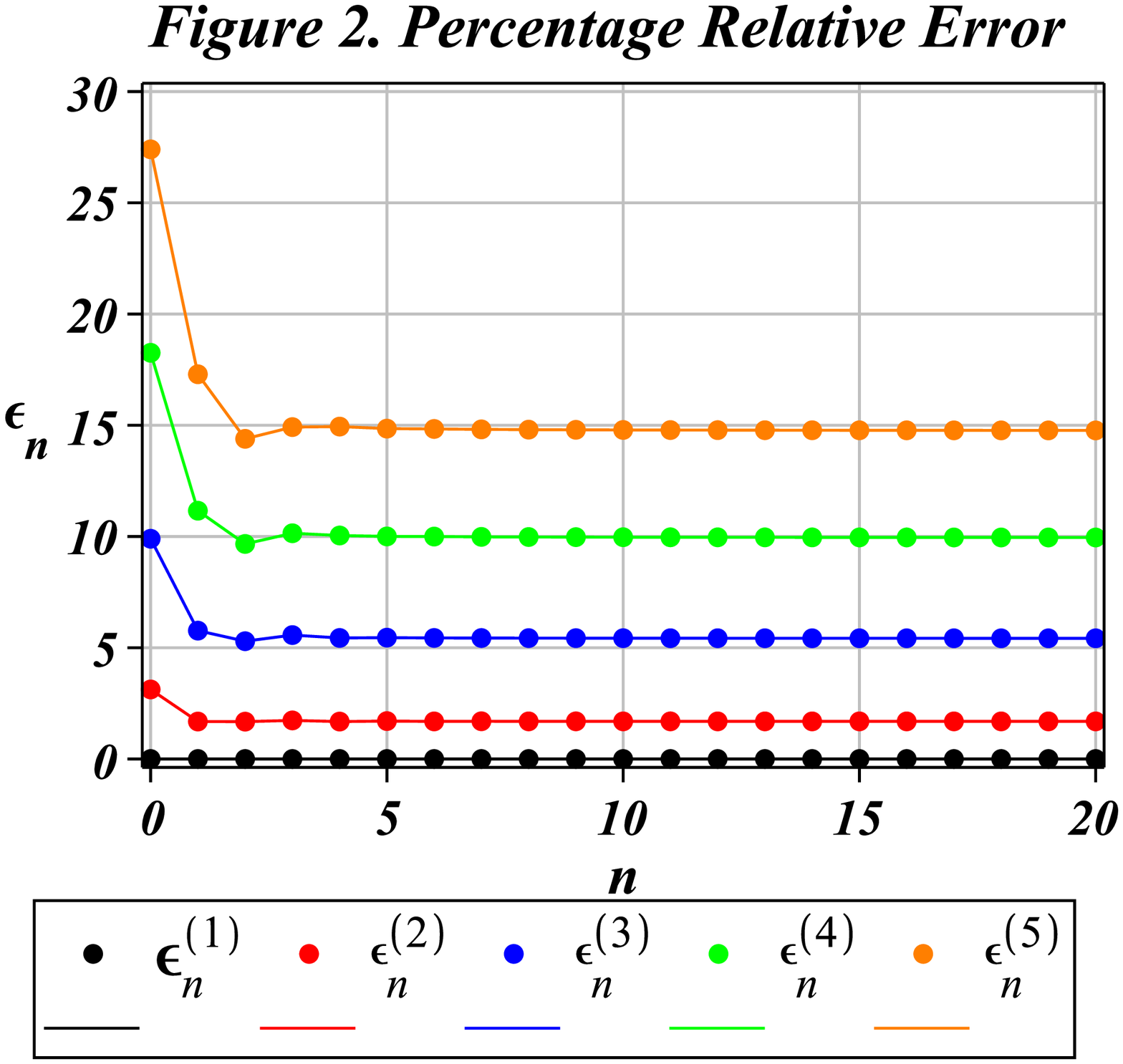}
\end{flushright}

\end{minipage}

\normalsize

\section{Discussion and perspectives}
The results obtained show that the procedure introduced in \cite{ans-free-P1} does not lead to satisfactory results when it is applied to $x^{2\kappa}$-type potentials, with $\kappa\geq 2$. In the other hand, we show that the relative errors in the energy eigenvalues do not depend on the coupling constant
 (coefficient of the monomial potential). This suggests that the relative errors for monomial potentials can be incorporate into the theory for establishing the upper bounds of the errors when the procedure is applied to polynomial potentials. 
To shed some light on this comment,  we are going to focus on the quartic anharmonic oscillator, which was treated in \cite{ans-free-P1}. 
The ans\"atze are given by (\ref{13}) and the approximate energy eigenvalues by (\ref{14}). Let’s start by observing that for small values 
of $\lambda$  ($\lambda \ll \omega^2/2$), the ans\"atze tend to those of the harmonic oscillator and for large values of the anharmonicity constant 
($\lambda \gg \omega^2/2$),  they tend to the ans\"atze of the pure quartic anharmonic oscillator (\ref{AHO4-3}).
We investigate the behavior of the energy eigenvalues with the values of $\lambda$. The first six energy eigenvalues, as functions of $\lambda$, 
are plotted in figure 3. The graphs corresponding to the  solutions obtained by computational numerical calculation are plotted with dashed black lines 
and the ans\"atze are plotted with solid lines 
(black for  $E_o^{ans}$, red for $E_1^{ans}$, blue for 	$E_2^{ans}$, green for $E_3^{ans}$, coral for $E_4^{ans}$ and violet for $E_5^{ans}$). 
We focus on each pair of curves for a given principal quantum number $n$ (ansatz-curve  vs numerical-curve).
 As can be observed for small $\lambda$-values, the such curves are overlapping (figure 3, left hand), 
 as $\lambda$ increases they  go grow apart, and when $\lambda \gg  \omega^2/2$ the spacing between them tends to a maximum constant value (figure 3, right hand). 
Then we studied the dependence of the  relative errors  with the $\lambda$-values.
 Figure 4 shows the percentage relative errors $\epsilon_n$ corresponding to the energy eigenvalues plotted in figure 3 (the same colors pattern was used). As can be observed, for each eigenstate with principal quantum number $n$, 
the $\epsilon_n$ depends on $\lambda$. For small values of  $\lambda$, $\epsilon_n$ is small and tends to zero when $\lambda$ tends to zero, 
which is consistent with the fact that the potential tends to that of the harmonic oscillator for which the procedure leads to the exact eigenvalues 
of energy  \cite{ans-free-P1}.
When $\lambda$ increases, $\epsilon_n$ increases tending to a maximum value $\epsilon_n^{max}$ for a theoretically infinite $\lambda$-value, which is consistent with the fact that when $\lambda$ is much greater than $\omega^2/2$, the harmonic part of the potential is negligible in front of the quartic contribution, and for the purely quartic potential the relative error is constant (depend on $n$ but not on the value of $\lambda$). 
These results suggest that the percentage relative errors  for the pure quartic potential (see tables 1) can be considered as the maximum values of the relative errors in the energy eigenvalues of the quartic anharmonic oscillator,
\ben  \epsilon_n=\epsilon_n(\lambda) \leq \epsilon_n^{max}\een

\vspace{0.2cm}

\small
\begin{center}
 {\bf Figure 3. First six energy eigenvalues as functions of $\lambda$-values ($\omega=1$). }
\includegraphics[width=15.50cm,height=6.0cm,angle=0]{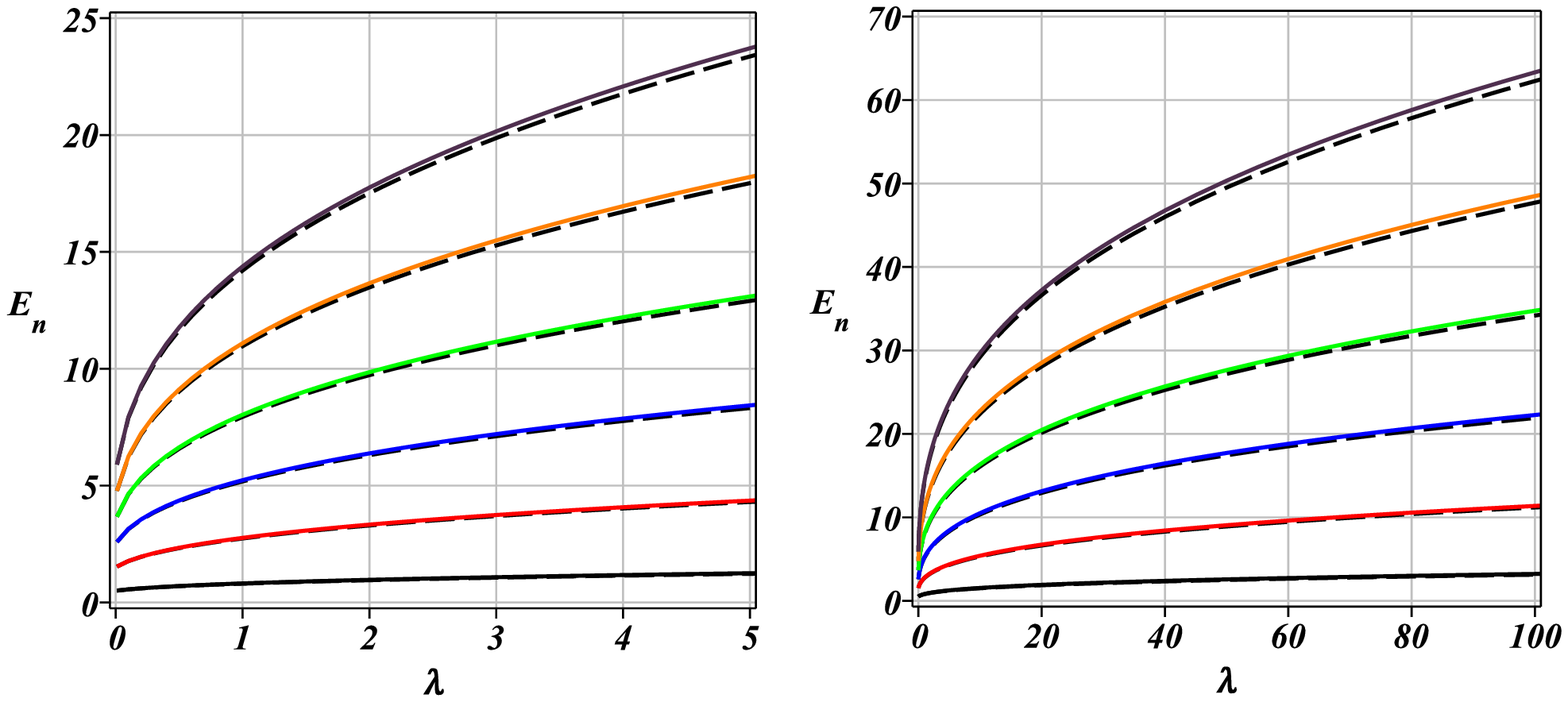}

\vspace{0.5cm}

 {\bf Figure 4. Percentage relative error $\epsilon_n$ as function of $\lambda$-values  ($\omega=1$). }
\includegraphics[width=15.0cm,height=6.cm,angle=0]{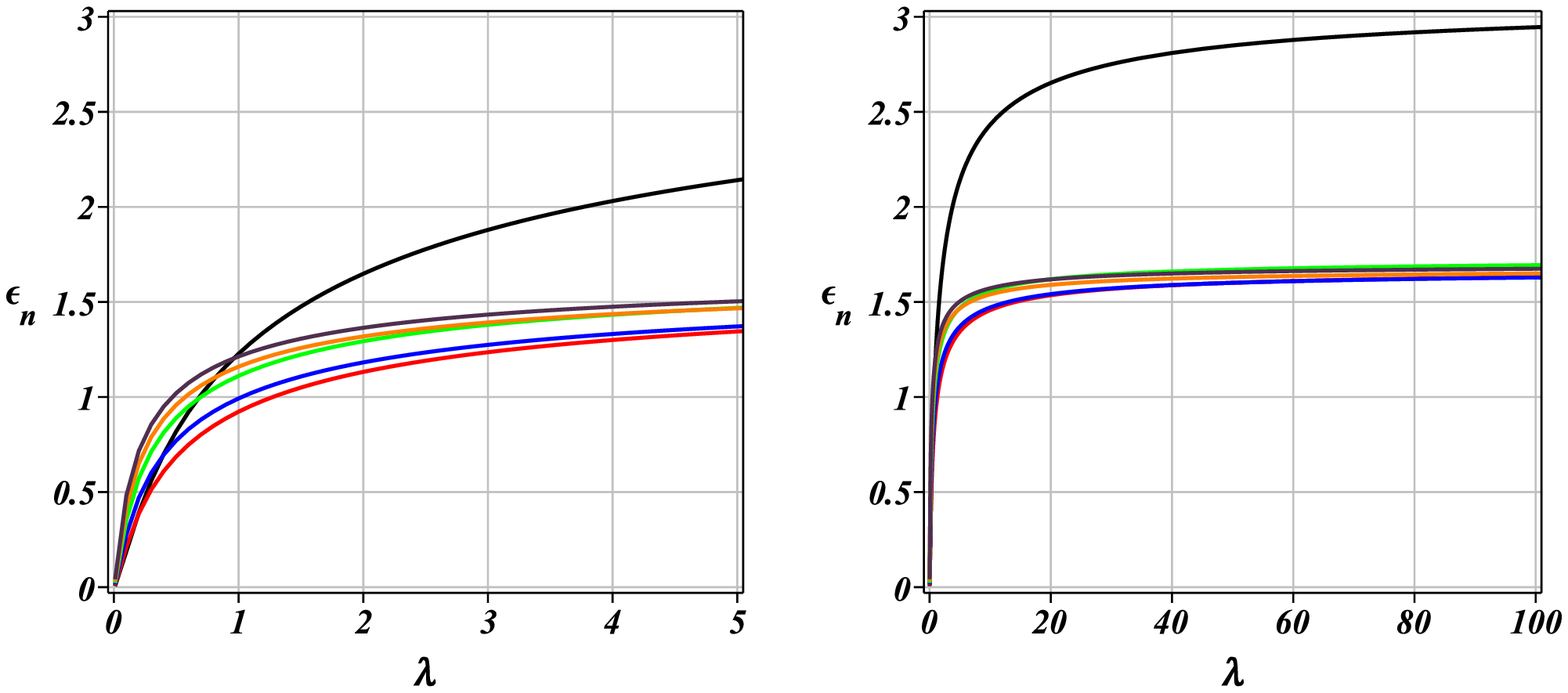}
\end{center}
\section{ Conclusions}
\nd The process derived from the virial theorem \cite{ans-free-P1} has been employed to obtain ans\"atze for  $x^{2\kappa}$-type potentials. We have found that the technique applied to high degree monomial potentials does not satisfactorily lead to the quantitative behavior of the system, since the relative errors increase considerably when the  degree of the monomial potential increases. 
In the other hand, since the ans\"atze preserve the  scaling properties of the Schr\"odinger equation,  the relative errors in the energy eigenvalues not depend on the  coefficient of the monomial potential. This property result helpful to estimate the error bounds when the technique is applied to polynomial potentials.
 We took a first step in this line, we study the behavior of the quartic harmonic oscillator with its parameters.
 The results suggest that, the procedure introduced in \cite{ans-free-P1}  provide satisfactory ans\"atze when it is applied to symmetric convex polynomial potentials, in which the largest of the polynomial coefficients is the one corresponding to the second degree term. Researches directed to establishing the error bounds as functions of the polynomial coefficients are in progress and will be reported elsewhere.

\vspace{0.2 cm}

\normalsize

\appendix
\renewcommand{\thesection}{\Alph{section}.\arabic{section}}
\renewcommand{\theequation}{A. \arabic{equation}}
\setcounter{equation}{0}
\begin{appendices}
\section*{Appendix: Scale transformation of the Schr\"odinger equation.}
\nd We describe here  the theory-changes under a scaling transformation for the SE given by
\be \label{scaling-1}
\left[-~\frac{1}{2}~\frac{d^2}{d x^2} +~ 
 \,\lambda\, x^{2\kappa}~\right]\psi_n(x) ~= ~ E_n~
\psi_n(x)~\ee  
\nd Under the scaling transform $ x \rightarrow v=\lambda^{1/[2(\kappa+1)]} x$ we get
\ben \label{scaling-1b}
\left[-~\frac{1}{2}~\lambda^{1/(\kappa + 1)}~\frac{d^2}{d {v}^{2}}+\lambda^{1-\kappa/(\kappa+1)}~v^{2\kappa}~\right]{\psi}_n^{sc}(v) ~= ~ { E}_n~ {\psi}_n^{sc}(v)~, \een 
 where ${{\psi}_n^{sc}}$ is the scaling transformed wave function of $\psi_n$.
Multiplying both terms of the above equation  by $\lambda^{-1/(\kappa + 1)}$ we obtain
\ben \label{scaling-2}
\left[-~\frac{1}{2}~\frac{d^2}{d {v}^{2}}+v^{2\kappa}~\right]{\psi}_n^{sc}(v) ~= ~ {\cal E}_n~ {\psi}_n^{sc}(v)~,\een 
 where
\ben \label{scaling-3}
 {\cal E}_n=\lambda^{-1/(\kappa+1)}E_n\een
Obviously, the eigenfunctions ${{\psi}_n^{sc}}$ of (\ref{scaling-2}) and their respective eigenvalues ${\cal E}_n$ are independent of $\lambda$-values.
\nd The inverse scaling transform $ v \rightarrow x=\lambda^{-1/[2(\kappa+1)]} v$ lead to
\ben \label{scaling-4}
{\psi}_n(x)=\alpha ~\psi_n^{sc}(\lambda^{1/[2(\kappa+1)]}~x)~, \een
 with $\alpha$ obtained requesting that both, the origin and the scaling transformed one, are normalized to unit,
\ben \label{scaling-4a} \int {\psi_n^2(x) dx} = \alpha^2 \int  {\left[\psi_n^{sc}(\lambda^{1/[2(\kappa+1)]}~x)\right]^2 dx}=
 {\alpha^2} \lambda^{-1/[2(\kappa+1)]}  \int \left[\psi_n^{sc}(v)\right]^2 ~dv \n\een
then
\ben \label{scaling-5} \int_{-\infty}^{~\infty} {\psi_n^2(x)~ dx} =   \int_{-\infty}^{~\infty} \left[\psi_n^{sc}(v)\right]^2 ~dv =1
 \hspace{0.4cm}\longrightarrow \hspace{0.4cm}\alpha~= \lambda^{1/[4(\kappa+1)]}~.\een
\nd Hence, the relation between the original and the scaling transform eigenfunctions  (\ref{scaling-4}) is given by 
\ben \label{scaling-6}
{\psi}_n(x)=\lambda^{1/[4(\kappa+1)]} ~\psi_n^{sc}(\lambda^{1/[2(\kappa+1)]}~x)~, \een
with the corresponding  relation between the energy eigenvalues (\ref{scaling-3}),
\ben \label{scaling-7}
 E_n=\lambda^{1/(\kappa+1)}{\cal E}_n\een
\end{appendices}

\linespread{1.0}
\small
%\bibliography{template}

\end{document}